\documentclass[footinbib,aip,apl,reprint]{revtex4-1}

\usepackage{graphicx}
\usepackage{amsmath}
\usepackage{mathtools}
\usepackage{comment}
\usepackage{bbm}
\usepackage{lipsum}
\usepackage{amssymb}
\usepackage{xcolor}
\definecolor{red}{rgb}{1,0.,0}
\newcommand{\fix}[1]{#1}

\begin{document}

\title{A broadband fiber-optic nonlinear interferometer}

\author{Joseph M. Lukens}
\email{lukensjm@ornl.gov}
\affiliation{Quantum Information Science Group, Computational Sciences and Engineering Division, Oak Ridge National Laboratory, Oak Ridge, Tennessee 37831, USA}

\author{Raphael C. Pooser}
\affiliation{Quantum Information Science Group, Computational Sciences and Engineering Division, Oak Ridge National Laboratory, Oak Ridge, Tennessee 37831, USA}
%\affiliation{Department of Physics, The University of Tennessee, Knoxville, Tennessee 37996, USA}

\author{Nicholas A. Peters}
\affiliation{Quantum Information Science Group, Computational Sciences and Engineering Division, Oak Ridge National Laboratory, Oak Ridge, Tennessee 37831, USA}
%\affiliation{Bredesen Center for Interdisciplinary Research and Graduate Education, The University of Tennessee, Knoxville, Tennessee 37996, USA}

\date{\today}

\begin{abstract}
We describe an all-fiber nonlinear interferometer based on four-wave mixing in highly nonlinear fiber. Our configuration realizes phase-sensitive interference with 97\% peak visibility and $>$90\% visibility over a broad 554~GHz optical band. By comparing the output noise power to the shot-noise level, we confirm noise cancellation at dark interference fringes, as required for quantum-enhanced sensitivity. Our device extends nonlinear interferometry to the important platform of highly nonlinear optical fiber, and could find application in a variety of fiber-based sensors.
\end{abstract}

\maketitle

Effects such as quantum back-action and sample damage limit the optical power and ultimately the sensitivity of interferometric sensors, making quantum methods the sole path for improvement in many situations.~\cite{Caves1981, Pooser2015a, Pooser2015b}  In particular, when such power-constrained sensors reach the shot-noise limit (SNL), the sensitivity can only be increased by incorporating squeezing or entanglement.~\cite{Giovannetti2004} \fix{In the most widespread configurations, quantum enhancement is realized by injecting quantum light into the unused port of a linear interferometer, methodology which has enabled improvements in the sensitivity of gravitational wave detectors~\cite{LIGO2011, Demkowicz2013} and even offers promise in sophisticated coupled interferometry as well.~\cite{Ruo2013,Ruo2015} Alternatively, one can modify the interferometer itself, supplanting the beam splitters of conventional interferometry with parametric amplifiers, thus realizing a so-called SU(1,1), or nonlinear, interferometer (NLI).~\cite{Yurke1986}} Bolstered by theoretical work predicting sub-SNL phase sensitivity with coherent-state inputs,~\cite{Plick2010,Ou2012} the field of nonlinear interferometry has expanded rapidly in recent years, with four-wave mixing (FWM) in $^{85}$Rb vapor proving a particularly popular platform.~\cite{Jing2011, Kong2013, Hudelist2014, Lukens2016, Anderson2017, Du2018} While free-space systems such as $^{85}$Rb and bulk $\chi^{(2)}$ crystals~\cite{Manceau2017b} have revealed the considerable potential of NLIs, they are not directly compatible with optical fiber, technology which underpins a variety of deployed sensors including gyroscopes, hydrophones, and current transducers.~\cite{Lee2003, Culshaw2008}

In this Letter, we outline work to bridge this gap, constructing and testing an SU(1,1) interferometer based on highly nonlinear fiber (HNLF). We observe 15~dB phase-insensitive gain, 97\% peak interference visibility, $>$90\% visibility over a wide $\sim$0.5~THz optical bandwidth, and noise cancellation at dark fringes. Our all-fiber NLI expands the reach of SU(1,1) interferometry, and could be incorporated into a range of fiber-optic sensors.

The core building block of any NLI is its nonlinear beam splitter: the parametric amplifier which enables probe and conjugate fields (alternatively, ``signal'' and ``idler'') to undergo a mixing operation described by the group SU(1,1).~\cite{Yurke1986} In the case of single-mode optical fiber, FWM performs just such a process, with two of the four interacting photons supplied by strong pump fields and the other two furnished by probe and conjugate modes themselves. Naturally occurring in silica and enhanced by the long lengths available in fiber optics, FWM can be boosted even further via HNLF possessing tight mode confinement and engineered dispersion profiles. Such FWM has been studied extensively in the context of phase-sensitive amplifiers (PSAs), which offer high gain, low loss, and noise figures in principle down to 0~dB.~\cite{Vasilyev2005,Tong2011,Tong2012,Slavik2012, Agarwal2014} Thus we enlist HNLF-based PSAs as nonlinear beam splitters in our NLI. After completing this work, we became aware of another approach for fiber-optic nonlinear interferometry based instead on dispersion-shifted fiber (DSF).~\cite{Liu2017} While sharing many of the practical advantages common to fiber optics in general, DSF has a lower nonlinear coefficient than HNLF and is also not commercially available. Therefore in making use of more ubiquitous HNLF, our approach has additional potential for improving the overall feasibility of fiber-optic NLIs.

\begin{figure*}[!t]
\includegraphics[width=6.5in]{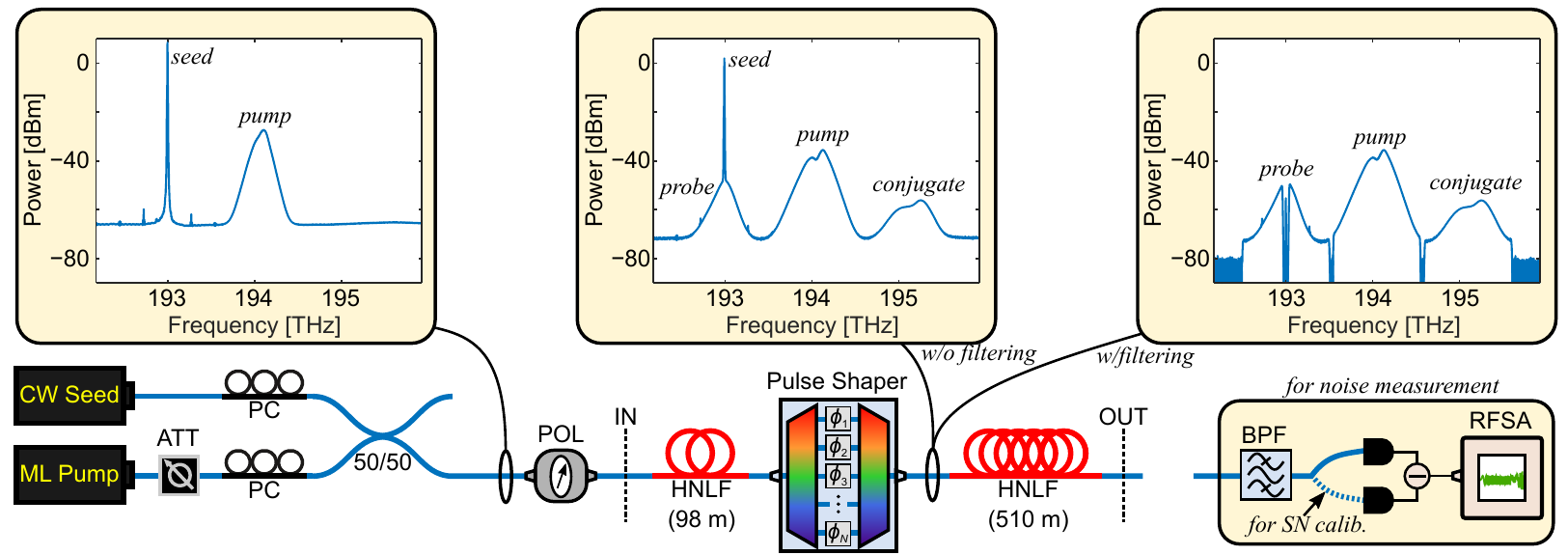}
\caption{Experimental setup, including measured optical spectra (at 2~GHz resolution) at various test points. ATT: variable attenuator, PC: polarization controller, POL: polarizer, BPF: bandpass filter, RFSA: radio-frequency spectrum analyzer.}
\label{fig1}
\end{figure*}

Figure~\ref{fig1} shows the experiment. The pump is a frequency-tunable, passively mode-locked laser (Laser-Femto) producing $\sim$5~ps pulses at 33.3~MHz; as seed, we use a tunable continuous-wave (CW) laser (Pure Photonics). In SU(1,1) interferometry, seeding the probe mode with such a coherent state enables sub-SNL phase uncertainty related to the parametric intensity gain $G$: $\Delta\phi = 1/(2G\sqrt{N})$, where $N$ is the number of sensing photons.~\cite{Ou2012} This enhancement holds at high fluxes, thus making it practical in application and much less sensitive to pump-induced noise, such as Raman scattering or residual spontaneous emission. As a consequence, all results are at room temperature, anticipating a reduced cost of size, weight, and power in deployed sensors.

\fix{We combine the two input fields with a 50/50 coupler and align their polarizations to  maximize power through an inline linear polarizer, in order to realize optimum FWM gain. The co-polarized fields then enter a 98~m long HNLF link (from OFS) possessing a zero-dispersion wavelength of 1542~nm (194.4~THz).} The pump is centered at 194.05~THz and attenuated initially to the point where self-phase modulation can just be observed, while the seed is set to 193.0~THz (chosen to optimize interference visibility below). The spectrum containing both the input seed and pump is shown in the top left of Fig.~\ref{fig1}; within the HNLF, pulsed probe and conjugate fields are generated via pump-degenerate FWM, as depicted in the top middle of Fig.~\ref{fig1}. To aid in system testing, we place a programmable optical pulse shaper (Finisar WaveShaper) after the first HNLF,  which compensates the chromatic dispersion of the system and applies an overall phase to each of the three frequency bands.~\cite{Weiner2011} The pulse shaper also defines 1~THz probe, pump, and conjugate passbands and filters out the residual CW seed light; the spectrum measured after filtering is depicted in Fig.~\ref{fig1} (upper right).

The common-spatial-mode nature of our interferometer ensures passive phase stability, with no need for active locking. \fix{It also preserves the co-polarized character of the interacting fields; even though fiber propagation modifies the overall polarization state in a complex manner, it does so identically for all three fields, obviating the need for additional polarization control before the second HNLF link.} However, unlike our intrinsically stable free-space NLI,~\cite{Lukens2016} the transmission between HNLF links is significantly lower here, due to the $\sim$4.5~dB insertion loss imparted by the pulse shaper. To maintain high gain in the second nonlinear beam splitter without separating and amplifying the pump, we employ a longer HNLF link. For example, if
we consider the simple approximation of monochromatic interacting beams and neglect pump depletion, the FWM gain can be written as
\begin{equation}
\label{e1}
G = 1 + \frac{\gamma^2 P^2}{\Delta\beta \left(\gamma P - \frac{\Delta\beta}{4} \right)} \sinh^2 \left\{\sqrt{\Delta\beta\left( \gamma P - \frac{\Delta\beta}{4} \right) } L   \right\},
\end{equation}
where $\Delta\beta$ is the wavevector mismatch between pump, probe, and conjugate fields, $\gamma$ the nonlinear coefficient, $P$ the pump power, and $L$ the fiber length.~\cite{Vasilyev2005} %Maximum gain results for $\Delta\beta = 2\gamma P$, and 
For two fiber links with $\{P_1,L_1\}$ and $\{P_2,L_2\}$, one can always find a set of frequencies (i.e., $\Delta\beta$ value) such that $G_2 \geq G_1$, as long as $L_2/L_1 \gtrsim P_1/P_2$. With $P_1/P_2\approx 3.5$ in our system, taking the second link at 510 m readily satisfies this condition. Admittedly, in the pulsed regime of the current experiments, this reasoning is only approximate, but it nonetheless represents a valuable starting point and allows us to realize comparable parametric gain values even with widely different pump powers.

%In nonlinear pulse propagation, the regime of operation is set roughly by the product of the nonlinear coefficient, input peak power, and fiber length ($\gamma P_\mathrm{peak} L$). Accordingly, by increasing the length of the second HNLF in proportion to the reduction in pump power (to 510~m in our case), we can attain similar nonlinear effects in the second link. Admittedly, we must stress that the gain's spectral dependence for the two fibers depends on power and length in a nontrivial way, so the two links do not really match across all wavelengths. However, this represents a valuable staring point and allows us to realize comparable parametric gain values even with widely different pump powers.

Incidentally, recent NLI studies have explored situations in which intentionally unbalancing the two nonlinear beam splitter gains could be beneficial in compensating loss.~\cite{Manceau2017b, Giese2017}  Therefore, our approach for varying fiber length to control gain under the restriction of limited pump power offers a useful added degree of freedom in NLI design---particularly in lossy environments---enhancing the value of our reported results and underscoring a natural strength of fiber-based NLIs.

\begin{figure}[!t]
\includegraphics[width=3.4in]{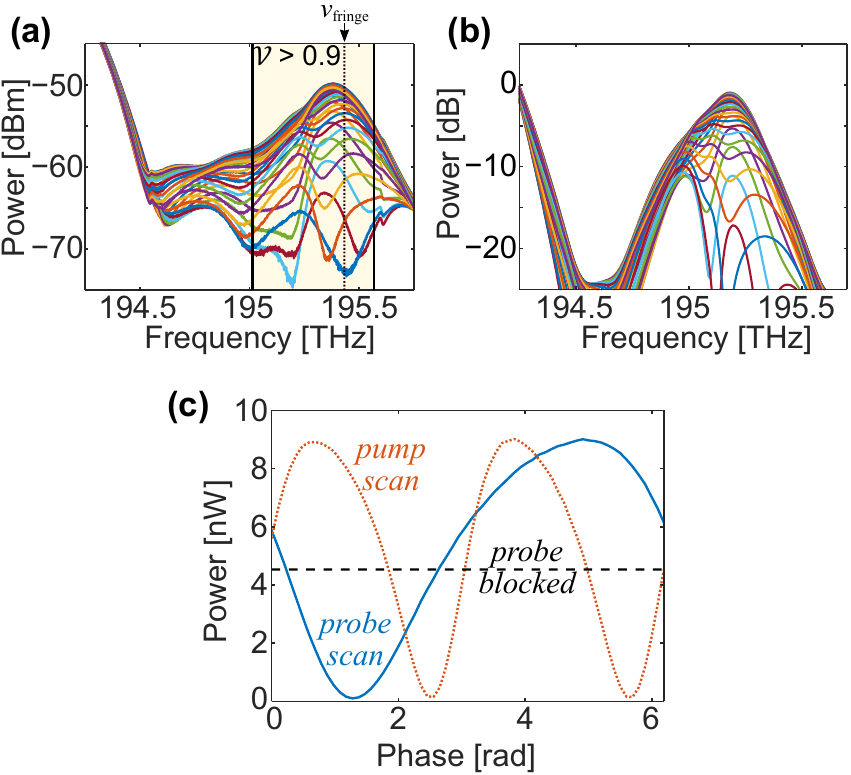}
\caption{(a) Measured conjugate spectra for 32 different probe phases. The shaded region denotes frequencies with visibilities exceeding $90\%$. (b) Simulation results for the same configurations as in (a). (c) Output power against phase at the maximum visibility point [$\nu_\mathrm{fringe}$ in (a)].}
\label{fig2}
\end{figure}

To look for phase-sensitive interference, we apply a uniform phase shift to the probe passband using the pulse shaper and measure the conjugate spectrum at the NLI output; Fig.~\ref{fig2}(a) plots the results of 32 different probe phases from 0 to $2\pi$. High-contrast interference is observed, with a 554~GHz bandwidth attaining visibilities in excess of 90\%, where at each frequency we define visibility in terms of maximum and minimum power: $\mathcal{V} = (P_\mathrm{max} - P_\mathrm{min})/(P_\mathrm{max} + P_\mathrm{min})$. This result highlights the broader bandwidth possible compared to more traditional atomic-based NLIs. Moreover, the experimental response qualitatively matches simulations obtained from the split-step Fourier method.~\cite{Agrawal2013,ssprop} In Fig.~\ref{fig2}(b), we plot numerical solutions of the nonlinear Schr\"{o}dinger equation using a combination of measurements and best estimates of our system parameters: \fix{for the average optical power at each stage, as well as the insertion loss of all components, we use experimental measurements; the HNLF parameters we take from the datasheet ($\gamma=9.3$~W$^{-1}$~km$^{-1}$, 1542~nm zero-dispersion wavelength, and 0.071~ps~nm$^{-2}$~km$^{-1}$ dispersion slope); and for the passively mode-locked pump, we assume a hyperbolic-secant shape with a full-width at half-maximum of 6~ps. Since the absolute power in Fig.~\ref{fig2}(a) depends heavily on the instrument response of our optical spectrum analyzer, for comparison purposes we plot the simulation results on an arbitrary relative scale covering 30~dB, for the same total range.}

\fix{Key experimental features such as the low-frequency shoulder and sharp dips in the high-contrast region of the spectrum are reproduced in simulation, indicating reasonable agreement with theory. The differences that are present, however, we believe result from uncertainty in our parameter estimates. Nonlinear effects are highly sensitive to the pump peak power, and we currently do not have access to tools (such as autocorrelation or spectral phase retrieval) to accurately determine the pulse shape; additionally, potential variations in dispersion through the HNLF could lead to nontrivial gain variations not incorporated into the simulation. We thus expect with further pulse and material characterization, greater theoretical and experimental congruity should be possible. Nonetheless, the current simulations do capture effects such as self-phase modulation, pump depletion, and differences in spectral gain profiles of the two fibers---typically neglected in basic NLI theory---which are likely responsible for the intricate features present in both Figs.~\ref{fig2}(a) and (b). Experimental investigation of these phenomena will also require more detailed optical and fiber characterization, which we intend to explore in future work.}

Figure~\ref{fig2}(c) shows the interference at frequency $\nu_\mathrm{fringe}= 195.433$~THz, marked on Fig.~\ref{fig2}(a), for scans of both the probe and pump phases. %The deviations from pure sinusoidal fringes are also in qualitative agreement with NLSE theory.
There is 15~dB of phase-insensitive gain and an additional 3~dB of phase-sensitive gain in the second parametric amplifier.~\cite{note1} As expected for single-pump FWM, the pump phase period is half that of the probe. The raw visibilities in Fig.~\ref{fig2}(b) are 0.976 and 0.967 for probe and pump phase scans, respectively. We emphasize that these visibilities are obtained from the average optical power directly, without any additional electronic filtering or locking to the pulse duration.%, thereby indicating low noise and good performance for a nonlinear optical system.

%\fix{By observing output optical power that oscillates with intra-interferometer phase, we confirm realization of an NLI: a device using parametric amplifiers to achieve interference. Yet for such an NLI to be useful---and offer advantages over linear interferometry---we not only need high-visibility interference, but also cancellation of correlated noise.}
Having shown the first major requirement of a successful NLI---output power that oscillates with intra-interferometer phase---we now study the second key enabling feature: cancellation of correlated noise. In the most straightforward case, quantum-enhanced sensitivity can be demonstrated via side-by-side comparison to a conventional interferometer.~\cite{Hudelist2014, Du2018} However, given the single spatial mode in our current setup, incorporating an equivalent linear interferometer would require substantial modifications. Nevertheless, we can characterize the noise scaling in our NLI and compare it against that required for quantum-enhanced sensitivity. In the ideal case (no loss, matched gain), the output conjugate field, with photon number operator $\hat{n}_c$, should possess super-Poissonian photon statistics at interference maxima, $\langle \Delta\hat{n}_c^2 \rangle > \langle\hat{n}_c \rangle$, and reduce to Poissonian statistics approaching the dark fringe, $\langle \Delta\hat{n}_c^2 \rangle = \langle\hat{n}_c \rangle$.~\cite{Lukens2016} Accordingly, a key signature of a properly functioning NLI is noise that varies with phase, exceeding the shot-noise level at maxima and approaching it at minima.

\begin{figure*}[!t]
\includegraphics[width=6.5in]{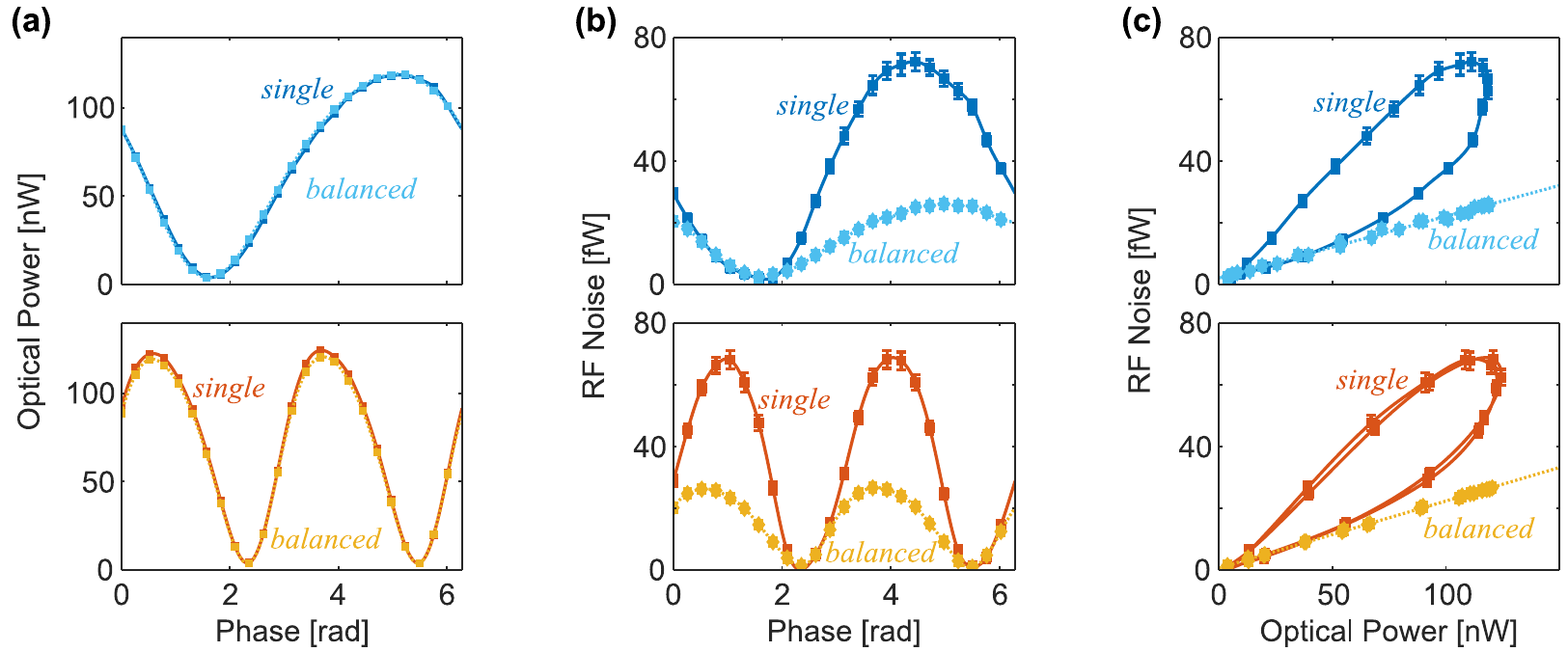}
\caption{Noise characterization of output conjugate field. The top row shows results from scanning the probe phase, the bottom row from scanning the pump. The labels ``single'' and ``balanced'' denote the two detector configurations. (a) Average optical power incident on both detectors. (b) RF noise measured on spectrum analyzer. (c) Noise plotted against optical power. Except for the linear fits in the balanced cases of (c), all curves are guides to the eye obtained from cubic spline interpolation.}
\label{fig3}
\end{figure*}

We employ the detection setup in the lower right inset of Fig.~\ref{fig1} for evaluating optical noise. To avoid potential complications from spectrally nonuniform gain, we filter out a $\sim$100~GHz optical band centered at 195.5~THz, and send it to a balanced pair of detectors (Thorlabs PDB570C). For measurement of the total noise, we direct all optical power to one detector, with the other blocked; to obtain the equivalent shot-noise level, we split the filtered conjugate field with a 50/50 fiber coupler and connect both paths to the detectors. Figure~\ref{fig3}(a) plots the total optical power incident on the detectors in each case, as the intra-NLI probe (top) and pump (bottom) phases are scanned;  the visibility for all four scans is 93\% within error. In Fig.~\ref{fig3}(b) we plot the radio-frequency (RF) noise level measured on the spectrum analyzer, after subtraction of the electronics floor. The frequency offset is 20~MHz, with resolution and video bandwidths of 100~kHz and 1~Hz, respectively; error bars are computed from the statistics of a 7.8~s analyzer trace. The noise does vary significantly between the two configurations: the single detector noise exceeds the balanced case by over a factor of three near bright fringes, but reduces to the balanced level at dark fringes.

To confirm that the balanced cases do represent the shot-noise level, we also plot the RF noise against the corresponding optical power. Figure~\ref{fig3}(c) verifies that the balanced detector noise scales as expected for shot noise; all data points fall on the linear best fit to within error. (The linear scaling down to low powers also establishes that the measurement is not limited by the electronics floor, but reflects true optical noise.) Interestingly, the single detector noise displays two distinct branches, reflecting different noise levels at the same average power, depending on the particular phase. We suspect that effects such as, e.g., gain saturation could be the cause of such asymmetric noise. Given the interplay between intra-NLI loss (which tends to decorrelate the optical noise) and spectral filtering (which likely eliminates a large portion of super-Poissonian noise), it is unclear how these absolute noise levels compare to theoretical predictions. Nonetheless, the dark-fringe reduction in optical noise relative to the shot-noise level confirms that significant cancellation is occurring in our system, and indicates the potential to realize quantum-enhanced sensitivity based on this design. \fix{Therefore, while do not perform full tomography to compare operation to ideal SU(1,1) Lie algebra, we do observe the two necessary features for useful SU(1,1) interferometry: phase-sensitive gain and correlated noise cancellation.}

%Nonetheless, the dark-fringe reduction in optical noise relative to the shot-noise level does confirm significant noise cancellation in our system, indicating the opportunity for quantum-enhanced sensors based on this HNLF design.}

Looking toward application in practical sensors, the flexibility realized by the commercial pulse shaper in the current setup should not be necessary for most specialized tasks, so that loss could be markedly reduced. For example, removing the pulse shaper from Fig.~\ref{fig1} gives a total loss from IN to OUT of only 1.8~dB, comparable to free-space NLIs based on $^{85}$Rb.~\cite{Du2018} If the quantity to be sensed imparts a spectrally selective phase shift, i.e., it modifies pump, probe, and conjugate fields differently, it should be possible even to utilize the current single-spatial mode structure in an installed system. In a more general situation, low-loss add-drop multiplexers could be inserted after the first HNLF link to couple the probe field to the device under test, and then return it to the second HNLF link for interference. Moreover, whereas here we consider three interacting pulsed modes to enhance peak power, single CW pumps have been shown to simultaneously amplify dozens of independent frequency pairs over multi-THz bandwidths,~\cite{Slavik2012} which in an NLI could facilitate spectrally resolved sensing of multiple quantities in parallel.

Finally, we note an interesting connection between our fiber-based NLI and previous work in fiber-optic parametric amplifiers. The sequence of seeded phase-insensitive amplification, loss, and phase-sensitive (de)amplification is physically equivalent to the so-called copier-PSA in classical communications.~\cite{Tong2011,Tong2012} However, the designed operating points differ significantly. To maximize output power, a copier-PSA should operate at a bright fringe, where correlations between probe and conjugate noise (produced in the copier stage) are phase-sensitively amplified, undesirably raising the noise figure of the PSA beyond its ideal value of 0~dB. Consequently, a large amount of loss between copier and PSA is beneficial to reduce correlated noise between the input fields and attain ideal copier-PSA performance.  As amplification in a communication channel may be applied after link transmission (loss), this requirement is reasonable in practice. On the other hand, an NLI realizes optimal sensitivity around the dark fringe, where probe and conjugate noise correlations are needed to produce the cancellation exploited for enhanced sensitivity, so loss is deleterious. In this way, our current work highlights a  useful, distinct, and complementary operating regime for fiber-optic parametric amplifiers.

%Finally, we note an interesting connection between our fiber-based NLI and previous work in fiber-optic parametric amplifiers. The sequence of seeded phase-insensitive amplification, loss, and phase-sensitive (de)amplification is physically equivalent to the copier-PSA in classical communications.~\cite{Tong2011,Tong2012} However, the designed operating points differ significantly. To maximize output power, a copier-PSA operates at a bright fringe, increasing the noise figure due to phase-sensitive amplification of correlated noise between the probe and conjugate [cf. Fig.~\ref{fig3}(b)]. Consequently, a large amount of loss between copier and PSA is desirable to reduce correlated noise.  As amplification in a communication channel may be applied after link transmission (loss), this requirement is practical. On the other hand, an NLI realizes optimal sensitivity around the dark fringe, where probe and conjugate noise correlations are needed to produce the cancellation exploited for enhanced sensitivity, so loss is deleterious. In this way, our current work highlights a  useful, distinct, and complementary operating regime for fiber-optic parametric amplifiers.

In conclusion, we have realized an all-fiber SU(1,1) interferometer, based on optical four-wave mixing in HNLF. We measure large gain, broad bandwidth, and high visibility, observing noise cancellation at dark fringes. Our NLI should have the potential to enable quantum-enhanced sensitivity in fiber-optic sensors.

\begin{acknowledgments}
This work was performed at Oak Ridge National Laboratory, operated by UT-Battelle for the U.S. Department of Energy under contract no. DE-AC05-00OR22725. Funding was provided by the Office of Naval Research (contract no. N0001418IP00047).
\end{acknowledgments}

%\bibliography{NLIref}
%merlin.mbs aipnum4-1.bst 2010-07-25 4.21a (PWD, AO, DPC) hacked
%Control: key (0)
%Control: author (8) initials jnrlst
%Control: editor formatted (1) identically to author
%Control: production of article title (-1) disabled
%Control: page (0) single
%Control: year (1) truncated
%Control: production of eprint (0) enabled
%

\end{document}